%% file: JCTCmain.tex
\documentclass[journal=jctcce,manuscript=article]{achemso}

\usepackage[version=3]{mhchem} 
\usepackage{lipsum}
\usepackage{accents}
\usepackage{mathrsfs}
\usepackage{physics}
\usepackage{braket}
\usepackage{mathtools}
\usepackage{amsfonts}
\usepackage{amsmath}
\usepackage{caption}
\usepackage{subcaption}
\usepackage{float}
\usepackage{geometry}
\usepackage{natbib}
\usepackage{setspace}
\usepackage{xkeyval}
\usepackage{multirow}
\usepackage{stackengine}
\usepackage{nicefrac}
\usepackage{titlesec}
\usepackage{booktabs}
\usepackage{varwidth}
\usepackage{hyperref}
\usepackage{enumitem}
\usepackage{esint}
\titleformat{\section}{\large\bfseries}{\thesection}{1em}{}
\usepackage[acronym]{glossaries}
\usepackage[usenames,dvipsnames]{color}
\DeclareMathAlphabet{\mathpzc}{OT1}{pzc}{m}{it}
\DeclareFontFamily{U}{mathx}{\hyphenchar\font45}
\DeclareFontShape{U}{mathx}{m}{n}{
      <5> <6> <7> <8> <9> <10>
      <10.95> <12> <14.4> <17.28> <20.74> <24.88>
      mathx10
      }{}
\DeclareSymbolFont{mathx}{U}{mathx}{m}{n}
\DeclareFontSubstitution{U}{mathx}{m}{n}
\DeclareMathAccent{\widecheck}{0}{mathx}{"71}
\DeclareMathAccent{\wideparen}{0}{mathx}{"75}
\DeclareMathAccent{\widebar}{0}{mathx}{"73}

\newcommand{\bc}{\boldsymbol{c}}

\newcommand{\bk}{\boldsymbol{k}}

\newcommand{\bu}{\boldsymbol{u}}

\newcommand{\bx}{\boldsymbol{\textbf{x}}}
\newcommand{\by}{\boldsymbol{\textbf{y}}}

\newcommand{\bF}{\boldsymbol{F}}
\newcommand{\bH}{\boldsymbol{\textbf{H}}}

\newcommand{\bK}{\boldsymbol{\textbf{K}}}

\newcommand{\bM}{\boldsymbol{\textbf{M}}}
\newcommand{\bL}{\boldsymbol{\textbf{L}}}

\newcommand{\bR}{\boldsymbol{\textbf{R}}}

\newcommand{\bphi}{\boldsymbol{\phi}}

\newcommand{\QE}{\texttt{QE}}
\newcommand{\mtA}{\texttt{CEF}}
\newcommand{\mtB}{\texttt{CPD}}
\newcommand{\LmtB}{\texttt{L-CPD}}

\newcommand{\DFTFE}{\texttt{DFT-FE}}

\newcommand{\cb}{\color{black}}

\newcommand{\cn}{\color{black}}

\newcommand{\comment}[1]{}
\author{Kartick Ramakrishnan}
\affiliation[CDS Indian Institute of Science, Bangalore]
{Department of Computational and Data Sciences, Indian Institute of Science, Bengaluru 560012, India}
\author{Gopalakrishnan Sai Gautam}
\affiliation[MT Indian Institute of Science, Bangalore]
{Department of Materials Engineering, Indian Institute of Science, Bengaluru 560012, India}
\email{saigautamg@iisc.ac.in}
\author{Phani Motamarri}
\email{phanim@iisc.ac.in}
\affiliation[CDS Indian Institute of Science, Bangalore]
{Department of Computational and Data Sciences, Indian Institute of Science, Bengaluru 560012, India}

\title
{ \Large Real-space methods for \emph{ab initio} modeling of surfaces and interfaces under external potential bias}

\abbreviations{}

\begin{document}





\begin{abstract}
Accurate \emph{ab initio} modelling of surfaces and interfaces, especially under an applied external potential bias, is important for describing and characterizing various phenomena that occur in electronic, catalytic, and energy storage devices. Leveraging the ability of real-space density functional theory (DFT) codes to accommodate generic boundary conditions, we introduce two methods for applying an external potential bias that can be suitable for modelling surfaces and interfaces. In the first method, an external constant electric field is applied by modifying the DFT Hamiltonian via introduction of an auxiliary linear potential while solving the electrostatic potential arising in DFT using a Poisson equation with zero-Neumann boundary conditions. The second method directly enforces the desired external potential bias by imposing constraints on the electrostatic potential, thereby naturally mimicking experimental conditions. We describe the underlying DFT governing equations for the two setups within the real-space formalism employing finite-element discretization. First, we validate the constant electric field setup within real-space finite-element DFT (DFT-FE) with an equivalent approach using plane-wave DFT (i.e., using periodic boundary conditions) on three representative benchmark systems, namely La-terminated Li$_7$La$_3$Zr$_2$O$_{12}$, GaAs (111), and Al FCC (111) slabs. Subsequently, we present a comprehensive evaluation of the two setups in terms of the average ground-state properties, such as  surface and adsorption energies{\cb, and introduce an approach to apply an external potential bias over a local region of a surface}. The methods developed in our work provide an attractive alternative to plane-wave DFT approaches in applying external potential bias that usually suffer from the periodic boundary conditions restrictions and poor scalability on parallel computing architectures. Our framework offers a robust approach for investigating surfaces and interfaces without any underlying assumptions or correction schemes while allowing for simulations of larger length scales than possible with plane-wave DFT.

\end{abstract}
\printglossary[type=\acronymtype]

\section{Introduction}
\input{intro}

\section{Methods}
\input{mathformulation.tex}

\section{Results}
\input{results.tex}

\section{Perspectives and concluding remarks}
\input{Discussion.tex}



\begin{acknowledgement}


\input{acknowledgement}
\end{acknowledgement}

\section*{Data availability}
All input and output calculation files that yielded the results presented in this work are available freely online in our \href{https://github.com/matrixlabiisc/AppliedPotentialBias_JCTC}{GitHub} repository.

\begin{suppinfo}
List of pseudopotentials used and comparison of differences in free energies, ionic forces, and dipole moments between \QE~ and \DFTFE~ in the three systems considered.
\end{suppinfo}

\bibliography{JCTC}
 \pagebreak


\end{document}

%% file: intro.tex
\emph{Ab initio} calculations of surfaces and interfaces provide a rigorous, atomistic-level understanding of the intrinsic properties that govern their behaviour in various applications. Accurate computation of surface energies, work functions, and the underlying ground-state electronic structures is critical for determining chemical reactivity, electronic phenomena, and catalytic adsorption. Typically, DFT simulations involving surfaces and interfaces are done using a 2-dimensional slab model of definite material thickness with a suitable vacuum layer employing periodic boundary conditions. This allows for detailed investigations into surface relaxation, surface reconstruction, defect formation, and diffusion mechanisms, which are pivotal in applications such as catalysis \cite{hammer_norskov_2000}, electronics \cite{zhang_yates_2012}, energy storage \cite{Peled_2017, xiao_ceder_2019}, and corrosion inhibition \cite{oguzie_wang_2011}. 

Plane-wave density functional theory (PW-DFT) is a widely used approach today for accurate electronic structure calculations employing pseudopotentials. The popularity stems from the systematic convergence offered by the plane wave basis set, which ensures spectral convergence in the computation of ground-state material properties\cite{Kresse_Abinitio, kaplan2023predictive}. However, this choice of the basis set restricts the simulation domains to be periodic. Furthermore, sufficient vacuum sizes or large cell sizes are required to minimise the interaction between their periodic images when computing ground state properties for molecules, nanoclusters, defective solids, slabs, and interfaces. Additionally, in the case of interfaces or surfaces with an intrinsic dipole moment, enforcing periodicity can lead to a spurious electric field, resulting in incorrect results and convergence issues \cite{Freysoldt_2020}. To mitigate these effects, larger symmetric slabs can be employed to eliminate net internal dipole moments. However, the requirement for large vacuums and larger slabs is computationally expensive, and the problem is compounded further due to the inferior scaling of plane wave codes on modern supercomputing architectures. 

To address the computational expense and convergence issues of surface/interface calculations, several strategies\cite{QEdipoleCorrection, NeugebauerDipoleSheet} have been proposed where an auxiliary linear potential is applied numerically to ensure the periodicity of electronic fields, often referred to as dipole correction schemes, which recover the original nature of the problem. \cb Other approaches include incorporating Coulomb truncation kernels\cite{ColoumbTruncation1, ColoumbTruncation2} that truncate the long-range Coulomb interactions between periodic images in plane-wave DFT codes.\cn~ In contrast, real-space methods such as finite difference methods\cite{parsec2006,rescu2016,sparc2017a,octopus2015,GPAW2010}, wavelet-based methods\cite{bigdft2008,bigdftgpu2009,bigdftlinearscaling2014} or finite-element (FE) methods\cite{white1989, tsuchida1995, tsuchida1996, pask1999, goedecker99, pask2005, bylaska, suryanarayana2010, motamarri2013, SCHAUER2013644, zhou2014, denis2016, Bikash2017, kanungo2019real, dftfe0.6, Das2019FastSystem} naturally accommodate generic boundary conditions and are systematically improvable while exhibiting excellent scalability on massively parallel computing architectures. In these methods, zero-Neumann boundary conditions on the electrostatic potential (i.e., the normal components of potential gradients are set to zero) can be imposed on the boundary parallel to the slab surface, with an additional constraint imposed to fix the electrostatic potential reference. Notably, for neutral slabs, zero-Neumann boundary conditions can eliminate the need for dipole correction schemes and large vacuum, improving accuracy and computational efficiency using real-space methods. 

Beyond analysing ground-state properties of material systems involving surfaces and interfaces, such as surface energies and work of adhesion, it is often necessary to investigate the other material parameters that control surface reactivity, diffusion mechanisms, and surface polarizability, to understand phenomena that occur at the application level. Such important material properties or parameters can be tuned by applying an external potential bias across the slab, i.e., by applying an external electric field. For example, applying an external potential bias enables the tuning of surface polarization to control ferroelectricity in nanoscale electronics \cite{C9NR05904K} and the manipulation of spin polarization for spintronic applications \cite{PhysRevLett.105.266806}. Additionally, external bias can regulate surface adsorption and modify chemical kinetics to enhance catalytic performance \cite{Pan2022, Che2018}. In battery systems, electrode$||$electrolyte interfaces experience significant electrostatic potential differences, which can alter ion migration pathways, ultimately impacting performance and efficiency.\cite{Famprikis_2019, Wang_Canepa_2022} Moreover, electrochemical impedance spectroscopy is a commonly used characterization technique in electrochemical devices that involves the application of an alternating potential bias across an interface to study the ionic and electronic transport mechanisms within the interface or material of interest.\cite{Chang_Park_2010} 

So far, PW-DFT calculations\cite{PhysRevB.64.125403, PhysRevB.63.205426, Galvez-Aranda_2019} have been used to provide theoretical insights into the effect of applying an external potential bias, where the bias is typically treated as a constant external electric field across the material system,\cite{SawToothResta, PhysRevB.64.125403, PhysRevB.73.165402, PhysRevLett.63.1617} which may not depict what is actually happening in a system, given that the electronic cloud within a solid will respond to any constant applied electric field. An alternative approach, based on Green’s functions\cite{Otani_ESM}, reformulates the electrostatic problem such that the computational domain is decoupled from periodic boundary conditions through an analytical form of Green’s functions for Poisson's equation for various boundary conditions. However, this method is restricted to boundary conditions for which an explicit analytical form of Green's function is available. Thus, it is important to develop calculation strategies to accurately model scenarios where an external potential bias is applied to a surface or an interface system, either during device operation or during characterization. \cb In this work, we focus on developing real-space strategies for applying an external electrostatic potential bias at a fixed number of electrons and demonstrate on materials systems with no net electron transfer. Note that extensions of real-space frameworks that incorporate constant chemical potentials allowing for the number of electrons to vary (i.e., grand canonical approaches\cite{gcdft}) are part of ongoing work.\cn


Here, we leverage the ability of real-space density functional theory (DFT) methodologies to accommodate generic boundary conditions to introduce two setups for applying an external potential bias across a slab system: (a) imposing a uniform constant external electric field (\mtA) and (b) directly applying a \cb constrained \cn~potential difference (\mtB). We introduce both setups using aperiodic boundary conditions in the DFT electrostatics problem. In the \mtA~setup, the external electric field is the tuning parameter that determines the resulting applied potential difference across the slab, as commonly done in plane-wave codes\cite{PhysRevB.64.125403, PhysRevB.63.205426, Galvez-Aranda_2019}.  To achieve a \mtA~in real-space DFT, we need to impose a constant external electric field in the non-periodic direction of the slab by modifying the DFT Hamiltonian, which is done by adding a sawtooth-shaped potential to the Kohn-Sham effective potential. Note that the linear segment of the sawtooth potential has a slope corresponding to the magnitude of the applied electric field. Further, the electrostatic problem involving the total charge density is solved by imposing zero-Neumann boundary conditions with a zero mean value constraint to fix the reference of the electrostatic potential. In the \mtB~setup, we directly control the electrostatic potential near the slab boundaries, providing a more natural representation of experimental setups where an explicit potential bias is applied. In the \mtB~approach, the underlying electrostatics problem corresponding to the total charge density is solved by imposing inhomogeneous boundary conditions that respect the external potential bias, which ensures that the potential bias across the slab is maintained during the self-consistent field iteration employed for solving the underlying DFT problem. \cb Further, we extend the \mtB~ approach by applying a potential difference over a locally constrained region of a surface (or the \LmtB~approach), which can't be trivially implemented in PW-DFT.\cn

We have adopted a finite-element (FE) methodology for solving the DFT problem in our current work. FE basis sets are systematically convergent and are compactly supported piecewise polynomial bases that can naturally accommodate generic boundary conditions. The locality of FE basis sets can exploit fine-grained parallelism on modern heterogeneous architectures while ensuring excellent scalability on distributed systems \cite{JPDC,Das2023}. Indeed, recent studies \cite{Das2023, Das2019FastSystem,dftfe0.6,dftfe1.0} have demonstrated that FE-based methods significantly outperform plane-wave approaches for norm-conserving pseudopotential DFT calculations, particularly for large systems to achieve a given accuracy of ground-state energy and forces. \cite{Bikash2017,ghosh2021}.
The open-source code \DFTFE~\cite{dftfe1.0, Das2023}, incorporates these features while leveraging scalable and efficient solvers for solving the Kohn-Sham equations. Additionally, the recently developed projector augmented wave method formalism within the FE framework (PAW-FE) \cite{pawfe} has demonstrated nearly a tenfold speedup over existing FE-based norm-conserving pseudopotential methods, thereby extending the length scales accessible to DFT computations. In this work, we implement both setups (\mtA~and \mtB) for applying an external potential bias within the \DFTFE~framework {\cb with norm-conserving pseudopotentials}, utilizing the advantages discussed earlier, thereby establishing a robust framework for large-scale simulations of surfaces and interfaces under an external potential bias.

We begin by benchmarking the \mtA~setup implemented in \DFTFE~with an equivalent approach \cite{SawToothResta, QEdipoleCorrection} used in plane wave codes. We consider three representative systems for our benchmarking, namely, La-terminated Li\textsubscript{7}La\textsubscript{3}Zr\textsubscript{2}O\textsubscript{12} (LLZO), GaAs(111), and Al FCC(111) slab, covering a diverse range of systems from polar to non-polar and from insulating to metallic, with applications in semiconductor devices, solid-state batteries, and catalysis. Subsequently, to examine the differences between the two setups (\mtA~vs.~\mtB) in \DFTFE, we plot the planar average electron density and planar average bare potential as a function of position along the non-periodic direction for the benchmark systems considered. Importantly, we observe that the bare potential for a given material system at the ground state is different between \mtA~and~\mtB, resulting in different ground-state solutions.  Finally, we compare the surface energy of (111) GaAs slab and La-terminated LLZO, and the adsorption energy of Na on the Al(111) surface as a function of the tuning parameters available in the \mtA~and \mtB~setups. \cb Also, we extend a comparison of the surface energies and dipole moments in the GaAs (111) surface between the \mtB~ and \LmtB~approaches. \cn  

The remainder of this article is structured as follows: Section 2 discusses the real-space formulation and FE discretization necessary for solving the Kohn-Sham ground-state problem. A detailed description of the two methods \cb (and extension to \LmtB) \cn of applying an external potential bias is presented. Section 3 presents a comprehensive benchmarking of the \mtA~setup against an equivalent approach used in plane-wave codes.  Following this, we demonstrate the differences between \mtA~and~\mtB~setups in \DFTFE~when applying an external potential bias across the slab and extend the comparison by evaluating the surface energy of GaAs(111) and La-terminated LLZO, as well as the adsorption energy of Na on Al(111). Finally, we discuss our observations,  outline future prospects arising from this work, and finish with a few concluding remarks.

%% file: mathformulation.tex
In this section, we outline the governing equations for determining the ground-state material properties involving slab models within the real-space formalism employed in this work. Subsequently, we examine different approaches for incorporating an external potential bias, detailing the modifications to the Hamiltonian, and the resulting energy and ionic forces expressions. Finally, we provide an overview of the FE formulation used in the current work to compute the ground-state properties of slabs under an applied potential bias.

\subsection{Governing equations and force expression in \texttt{DFT-FE}}
The ground-state properties of a slab comprising of $N_a$ nuclei and $N_e$ electrons in a representative supercell within the norm-conserving pseudopotential formalism are governed by the following Kohn-Sham density functional theory (DFT) energy functional,\cite{parr1979local,martin2020electronic}
\begin{equation}
    E\left[ \left\{ \psi_n \right\},\left\{\bR^a \right\} \right] = \min_{\{\psi_n\} \in \chi(\Omega_p)}{\left\{ T_\text{s} + E_{\text{xc}} +  E_{\text{el}} + E_{\text{psp}} \right\}}
    \label{eq: KS energy}
\end{equation}
where $\left\{ \psi_n \right\}$ denotes the single-electron  wavefunctions satisfying the orthonormality condition $\braket{\psi_i|\psi_j}=\delta_{ij}$ with $1 \leq n \leq N$ where $N \geq \frac{N_{e}}{2}$, and $\left\{\bR^a \right\}$ signifies the position vectors of the $N_a$ nuclei. We note that $\chi(\Omega_p)$ denotes an appropriate function space in which the single-electron wavefunctions lie, with $\Omega_p$ representing the 2D periodic slab domain. We focus here on the spin-unpolarized formulation for clarity and notational convenience, while the extension to the spin-polarized framework is straightforward. 

\cb The term $T_\text{s}$ in Equation~\eqref{eq: KS energy} represents the kinetic energy of the non-interacting electrons, while $E_{\text{xc}}$  represents the exchange-correlation energy that accounts for the quantum mechanical many-body effects, and are given by
\begin{equation}
    T_{\text{s}}[\{\psi_n\}]= 2\sum_n{f_n\int{\frac{1}{2}|\nabla \psi_n(\bx)|^2 d\bx}},\;\;\;\; E_{\text{xc}}[\rho(\bx)] = \int{\epsilon_{\text{xc}}\left[\rho(\bx),\nabla \rho(\bx)\right]d\bx} \label{eq:Tsexc}
\end{equation}
where,  the generalised gradient approximation\cite{Langreth19831809, martin2020electronic,GGA1997} (GGA) has been adopted for the exchange-correlation contribution. Further, the electron density ($\rho(\bx)$) and its gradient($\nabla \rho(\bx)$) in Equation~\eqref{eq:Tsexc} are computed as,\cn
\begin{equation}
    \rho(\bx) = 2\sum_n^{N}{f_n|\psi_n(\bx)|^2};\;\; \nabla \rho(\bx) = 2\sum_n^N{f_n\bigg(\psi_n^*(\bx)\nabla\psi_n(\bx)+\psi_n(\bx)\nabla\psi_n^*(\bx)\bigg)}
\end{equation}
with $\bx$ denoting the spatial coordinate, and $f_n$ is the occupation number corresponding to the electronic wavefunction indexed by `$n$' in the above. Furthermore, $E_{\text{el}}$ in Equation~{\eqref{eq: KS energy}} represents the classical electrostatics energy computed as 
\begin{equation}
    E_{\text{el}}[\rho(\bx),\{\bR^a\}] = \cb \max_{\hat{\phi} \in \kappa(\Omega_p)}{\left\{\int_{\Omega_p}{(\rho(\bx)+b(\bx))\hat{\phi}(\bx)d\bx}-\frac{1}{8\pi}\int_{\Omega_p}{|\nabla \hat{\phi}(\bx)|^2d\bx}\right\}} \cn - \sum_a{E^a_{\text{self}}} 
\label{eq: electrostatics Energy}
    \end{equation}
where $\hat{\phi}(\bx)$ denotes the trial function for the electrostatic potential due to the total charge density $(\rho(\bx)+b(\bx))$ and belongs to a suitable function space $\kappa(\Omega_p)$. \cb Additionally, $E^a_{\text{self}}$ in Equation~\ref{eq: electrostatics Energy} represents the self-energy associated with an atom-centered smeared charge density $b^a_{\text{sm}}(\bx)$. $E^a_{\text{self}}$ arises from introducing the atom-centered smeared charges in the local real-space electrostatics reformulation, which leads to an additional atom-centered potential, $V^a_{\text{sm}}(\bx-\bR^a)$ (see Equation 17 in our previous work\cite{dftfe1.0})\cn. Moreover, the electrostatic potential $\phi(\bx)$ generated by the total charge density ($\rho(\bx)+b(\bx)$), where $b(\bx) = \sum_a{b_{\text{sm}}^a(\bx-\bR^a)}$, is obtained by solving the following Poisson's equation,
\begin{equation}
    -\frac{1}{4\pi}\nabla^2 \phi(\bx) = \rho(\bx) + b(\bx),
    \label{eqn: Poissons equation}
\end{equation}
\cb that denotes the Euler-Lagrange equation associated with the maximization problem in Equation~\eqref{eq: electrostatics Energy}\cn. Finally, $E_{\text{psp}}[\{\psi_n\},\{\bR^a\}]$ in Equation~{\eqref{eq: KS energy}}~represents the pseudopotential energy contribution, which is written as the sum of local and nonlocal contributions, i.e., $E_{\text{psp}}[\{\psi_n\},\{\bR^a\}] = E_{\text{loc}}[\rho(\bx)] + E_{\text{nloc}}[\{\psi_n\},\{\bR^a\}]$, where the local pseudopotential, $E_{\text{loc}}$, is expressed as,
\begin{equation}
   E_{\text{loc}}[\rho(\bx)]  = \int_{\Omega_p}{\big(V_{\text{loc}}(\bx) - V_{\text{sm}}(\bx)\big)\rho(\bx) d\bx}
   \label{eq: local pseudopotential}
\end{equation}
In the above equation, $V_{\text{loc}}(\bx)$ is the sum of atom-dependent local pseudopotentials, i.e., $V_{\text{loc}}(\bx) = \sum_a{V_{\text{loc}}^a}(\bx-\bR^a)$, \cb while $ V_{\text{sm}}(\bx) = \sum_a{ V_{\text{sm}}^a}(\bx-\bR^a)$. \cn
Since the electrostatic energy computed from Equation~\eqref{eq: electrostatics Energy} involves the electrostatic potential $\phi(\bx)$ due to the electron charge and smeared charge density, i.e., $\rho(\bx)+b(\bx)$, the energy contribution given by $E_{\text{loc}}[\rho(\bx)]$ term not only includes the contribution from  $V_{\text{loc}}(\bx)$ but is adjusted to exclude the contribution of $V_{\text{sm}}(\bx)$ arising from the smeared charge density $b(\bx)$.
Additionally, the non-local pseudopotential energy contribution, $E_{\text{nloc}}$, for optimized norm-conserving (or ONCV) pseuodpotentials\cite{oncv2013} is given by
\begin{equation}
    E_{\text{nloc}}[\{\psi_n\},\{\bR^a\}] = 2\sum_n{f_n{\int_{\Omega_p}{\int_{\mathbb{R}^3}{\psi_n^*(\bx)V_{\text{nloc}}(\bx,\by) \psi_n(\by)d\by}d\bx}}}
\end{equation}
where the action of  $V_{\text{nloc}}(\bx,\by)$ on electronic wavefunction is given by
\begin{equation}
    \int_{\mathbb{R}^3}{V_{\text{nloc}}(\bx,\by)\psi_n(\by)d\by} := \sum_a^{N_a}{\sum_{\beta}{p^a_{\beta}(\bx-\bR^a)D^a_{\beta}\int_{\mathbb{R}^3}{p^a_{\beta}(\by-\bR^a)\psi_n(\by)d\by}}}
    \label{eq: V_nloc}
\end{equation}
The composite index $\beta=\{ n,l,m \}$ in Equation~{\eqref{eq: V_nloc}} is such that $l$ and $m$ denote the orbital and magnetic angular momentum index, respectively. Furthermore, $p_{\beta}^a(\bx-\bR^a)$ indicates the atom-centred projector of index $\beta$ while $D^a_{\beta}$ represents the pseudopotential coupling coefficients. Finally, the problem of determining the ground-state properties for  given positions of nuclei ($\{\bR^a\}$) is determined by solving the following variational problem,
\cb \begin{multline}
   E_{\text{GS}}\left[ \left\{ \psi_n \right\},\left\{\bR^a \right\} \right] = \min_{\{\psi_n\}\in \chi(\Omega_p)}{\;\max_{\hat{\phi}\in \kappa(\Omega_p)}{\bigg\{T_s[\left\{ \psi_n \right\}] -\frac{1}{8\pi}\int_{\Omega_p}{|\nabla \hat{\phi}(\bx)|^2d\bx} + \int_{\Omega_p}{(\rho(\bx)+b(\bx))\hat{\phi}(\bx)d\bx}}} \\
    {{-\sum_a^{N_a}{E^a_{\text{self}}}+E_{\text{xc}}[\rho(\bx)]+ E_{\text{loc}}[\rho(\bx)] + E_{\text{nloc}}[\{\psi_n\},\{\bR^a\}] \bigg\} }}
    \label{eq: KS DFT variational problem}  
\end{multline}\cn

The Euler-Lagrange equation corresponding to the minimization of the energy functional in Equation~\eqref{eq: KS DFT variational problem} subject to the orthonormality constraint on the single-electron wavefunctions ($\int{\psi_i^*(\bx)\psi_j(\bx)d\bx}=\delta_{ij}$) leads to the  Hermitian eigenvalue problem $\mathcal{H}\psi_i = \varepsilon_i\psi_i$ that needs to be solved for the smallest $N \geq N_e$ eigenpairs $\{\varepsilon_i,\psi_i\}$ of the Hamiltonian operator $\mathcal{H}$. In turn, $\mathcal{H}$ is decomposed as $\mathcal{H} = \mathcal{H}_{\text{loc}}+\mathcal{H}_{\text{nloc}}$, where $\mathcal{H}_{\text{loc}}$ is defined to be,
\begin{equation}
    \mathcal{H}_{\text{loc}} = \bigg[ -\frac{1}{2}\nabla^2 + V_{\text{eff}}(\bx) \bigg],\;\; \text{with}\;\;\; V_{\text{eff}}(\bx) := \bigg( \frac{\delta E_{\text{xc}}}{\delta \rho(\bx)}[\rho(\bx),\nabla \rho(\bx)] +  \widetilde{\phi}(\bx) \bigg)
    \label{eq: Veff and Hloc}
    \end{equation}
  with $\widetilde{\phi}(\bx) := \phi(\bx) + V_{\text{loc}}(\bx) - V_{\text{sm}}(\bx)$, henceforth referred to as total electrostatic potential which includes the electron-electron and electron-nuclear interactions.  
 Additionally, the action of $\mathcal{H}_{\text{nloc}}$ on wavefunction$\psi_n(\bx)$ is defined as,
 \begin{equation}
   \mathcal{H}_{\text{nloc}}\psi_n := \sum_a^{N^a}\int_{\mathbb{R}^3}{ V^a_{\text{nloc}}(\bx,\by)\psi_n(\by) d\by}  
 \end{equation}
 
When dealing with periodic crystals, 2D slabs or surfaces, it is computationally efficient to invoke Bloch's theorem\cite{ashcroft2022solid,martin2020electronic} along the periodic directions and instead of solving the problem on large periodic supercells, we solve the problem on smaller unit-cells with periodic boundary conditions. \cb Using Bloch’s theorem, the electronic wavefunction can be expressed as $\psi_{n\bk}(\bx) = e^{i\bk\cdot\bx}u_{n\bk}(\bx)$, where $i = \sqrt{-1}$ and $u_{n\bk}(\bx)$ is a lattice-periodic function satisfying $u_{n\bk}(\bx + \bL_r) = u_{n\bk}(\bx)$ for all reciprocal lattice vectors $\bk$ in the first Brillouin zone and for all lattice vectors $\bL_r$ in the periodic directions. \cn To this end, the governing equations involving Bloch wavefunctions to be solved for determining the ground-state properties are given as follows:
\cb \begin{gather}   
    \mathcal{H}^{\bk}u_{n\bk} = \varepsilon_{n\bk}u_{n\bk} \;\; \text{with}\;\;
     \mathcal{H}^{\bk} := \mathcal{H}^{\bk}_{\text{loc}} + \mathcal{H}^{\bk}_{\text{nloc}}  \nonumber \\
   -\frac{1}{4\pi} \nabla^2 V^a_{\text{sm}} = b^a_{\text{sm}}(\bx-\bR^a), \;\;       -\frac{1}{4\pi}\nabla^2 \phi(\bx) = \rho(\bx) + b(\bx)
       \label{eq: Governing equations}
     \end{gather}
  where, $\mathcal{H}^{\bk}_{\text{loc}}$ and the action of $\mathcal{H}^{\bk}_{\text{nloc}}$ on a wavefunction is given by  
 \begin{gather}   
      \mathcal{H}^{\bk}_{\text{loc}}  := \bigg[-\frac{1}{2}\nabla^2 -i\bk\cdot\nabla + \frac{1}{2}|\bk|^2 + V_{\text{eff}}(\bx) \bigg] ,\;\;
      V_{\text{eff}}(\bx) = \bigg[ \frac{\delta E_{\text{xc}}[\rho(\bx),\nabla \rho(\bx)]}{\delta \rho(\bx)} + \widetilde{\phi}(\bx) \bigg] \nonumber \\
   \widetilde{\phi}(\bx) = \phi(\bx)+ \left(V_{\text{loc}}(\bx) - V_{\text{sm}}(\bx)\right )\nonumber \\
   V_{\text{loc}}(\bx) =  \sum_r{\sum_{a\in \Omega_p}{V^a_{\text{loc}}(\bx-\bR^a+\bL_r)}} ,\;\;V_{\text{sm}}(\bx) = \sum_r{\sum_{a\in \Omega_p}{V^a_{\text{sm}}(\bx-\bR^a+\bL_r)}} \nonumber \\   
   \rho(\bx) = 2\sum_n{\fint_{BZ}{f_{n\bk}|u_{n\bk}(\bx)|^2d\bk}}, \;\;\nabla \rho(\bx) = 2\sum_n{\fint_{BZ}{f_{n\bk}(u^*_{n\bk}(\bx)\nabla u_{n\bk}(\bx)+ u_{n\bk}(\bx)\nabla u^*_{n\bk}(\bx))d\bk}} \nonumber \\ 
      \mathcal{H}^{\bk}_{\text{nloc}}u_{n\bk} := \sum_{a \in \Omega_p}{\sum_{\beta}{\sum_{r}e^{-i\bk\cdot(\bx-\bL_r)}p^a_{\beta}(\bx-\bR^a-\bL_r) D^a_{\beta}\int_{\Omega_p}{\sum_{r'}{ e^{i\bk\cdot(\by-\bL_{r'})}p^a_{\beta}(\by-\bR^a-\bL_{r'})u_{n\bk}(\by) d\by}} }} 
       \label{eq: Governing equations1}
     \end{gather}   \cn  
where $\fint_{BZ}$ denotes the volume average of the integral over the first Brillouin zone (BZ) corresponding to the periodic unit cell $\Omega_p$. A detailed discussion on computing ion forces and cell stresses can be found in our previous work \cite{dftfe1.0, configurationalForces}. For completeness, we mention here the expressions for the ionic force in the norm-conserving pseudopotential formulation:
 \small{
 \begin{align}
   &\bF^{a} = -\frac{d E_{\text{GS}}}{d \bR^a} = F^a_{\text{loc}} + F^a_{\text{nloc}} + F^{a^*}_{\text{nloc}} \label{eq: Force expression} \\
   &\bF^a_{\text{loc}} = -\int_{\Omega_p}{\nabla \rho(\bx)\big(V^a_{\text{loc}}(\bx-\bR^a)-V^a_{\text{sm}}(\bx-\bR^a)\bigr)d\bx} -\frac{1}{2}\int_{\Omega_p}{b^a_{\text{sm}}(\bx-\bR^a)\nabla\phi(\bx)d\bx} \nonumber
   \end{align}
   \begin{multline*}
   \bF^a_{\text{nloc}} = 2\sum_{n=1}^N{\fint_{BZ}{f_{n,\bk}\biggl[\int_{\Omega_p}{\sum_{\beta}{\sum_{r}{e^{-i\bk\cdot(\bx-\bL_r)}p^a_{\beta}(\bx-\bR^a-\bL_r)d\bx D^a_{\beta}}}\int_{\Omega_p}{\sum_{r'}{e^{i\bk\cdot(\by-L_{r'})}}}}}} \\
    {{{{{p^a_{\beta}(\by-\bR^a-\bL_{r'})(-\nabla u_{n\bk}(\by) + i\bk u_{n\bk}(\by))} d\by}}\biggr]d\bk}} 
 \end{multline*}
 }

\subsection{Applying an external potential bias}
First, we consider the effect of a constant external electric field (or \mtA), which is analogous to the sawtooth method with dipole correction employed in plane-wave codes\cite{QEdipoleCorrection,SawToothResta}. Second, we examine the case where the classical electrostatic potential ($\phi(\bx)$) conforms to the external potential bias through the boundary conditions imposed on $\phi(\bx)$ in the Poisson equation (Equation~\eqref{eqn: Poissons equation}), referred to as the constrained potential difference (or \mtB) setup. 

\subsubsection{\mtA~setup}\label{sec: constant electric field}
\begin{figure}[htp]
\centering
    \includegraphics[width=\textwidth]{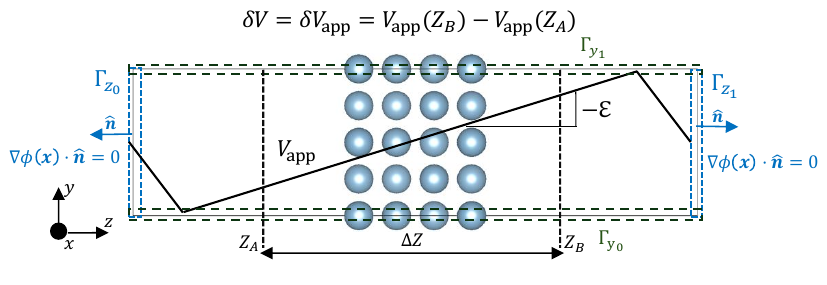}
\caption{\textbf{\mtA}$:$ The boundary conditions for the  electrostatic potential (${\phi}(\bx)$) are periodic along the $x$ and $y$ directions. $\Gamma_{y_0}$, $\Gamma_{y1}$, $\Gamma_{x_0}$(not shown), and $\Gamma_{x_1}$(not shown) denote the periodic boundaries.  Zero-Neumann conditions are applied on the non-periodic boundaries $\Gamma_{z_0}$, $\Gamma_{z_1}$. The constant electric field is applied by introducing the saw-tooth potential $V_{\text{app}}$. The electric field ($\mathcal{E}$) is determined  as $-\mathcal{E}\Delta Z = \delta V$).}
        \label{fig:CEF}
\end{figure}

The typical approach to imposing an external electric field on slabs is to employ a sawtooth potential\cite{SawToothResta}, which is the method of choice for slabs in Quantum Espresso (\QE\cite{qe}) and one of the approaches used in our work as well. An alternative approach to applying a constant electric field is by introducing a dipole sheet in the vacuum region\cite{NeugebauerDipoleSheet, PhysRevB.64.125403}. In the current \mtA~ setup, the effective potential ($V_{\text{eff}}(\bx)$) in Equation~\eqref{eq: Veff and Hloc} is modified to take the form,
\begin{equation}
  V_{\text{eff}}(\bx) = \bigg( \frac{\delta E_{\text{xc}}}{\delta \rho(\bx)}[\rho(\bx),\nabla \rho(\bx)] +  \widetilde{\phi}(\bx) + V_\text{app}(\bx) \bigg) 
\end{equation}
where $V_{\text{app}}(\bx)$ is the linear periodic potential across the material system as indicated in Figure~\ref{fig:CEF}. The slope of $V_{\text{app}}$ dictates the magnitude of the electric field, $-\mathcal{E} =\frac{d V_{\text{app}}}{dz}$ \cb, and ensures a straightforward benchmarking with plane-wave codes \cn. We design the profile of $V_{\text{app}}(\bx)$ to be a sawtooth function, similar to plane-wave codes, with the maximum and the minimum values located close to the simulation cell boundaries, as shown in Figure~\ref{fig:CEF}. \cb Additionally the sawtooth form acts as a constraining potential, preventing electron density leakage into the low-potential regions, allowing for better self-consistent field convergence, especially at higher magnitudes of $\mathcal{E}$.\cn~Unlike plane-wave basis, the electrostatic potential (${\phi}(\bx)$) in \DFTFE~need not be fully periodic and we impose semi-periodic boundary conditions on ${\phi}(\bx)$ to simulate neutral slabs. Specifically, as displayed in Figure~\ref{fig:CEF}, we impose periodic boundary conditions in the $x,y$-planar directions and a zero-Neumann boundary condition on the boundaries parallel to the slab surface while solving for $\phi(\bx)$ using the Poisson equation (see Equation~\eqref{eqn: Poissons equation}). Additionally, we apply a zero mean-value constraint, $\int_{\Omega_p}{{\phi}(\bx)d\bx} = 0$, to fix the reference potential and remove the arbitrary constant offset. 

Finally, we summarize the governing equation and boundary conditions to determine $\phi(\bx)$ in the \mtA~setup as:
\begin{equation}
\left\{
\begin{aligned}
&-\frac{1}{4\pi}\nabla^2 \phi(\bx) = \rho(\bx) + b(\bx), \quad \bx \in \Omega_p,\\[1ex]
&\phi(\bx)\Big|_{\Gamma_{x_0}} = \phi(\bx)\Big|_{\Gamma_{x_1}}, \quad \phi(\bx)\Big|_{\Gamma_{y_0}}= \phi(\bx)\Big|_{\Gamma_{y_1}},\\[1ex]
&\nabla \phi(\bx) \cdot \hat{\boldsymbol{n}}\Big|_{\Gamma_{z_0}} = 0, \quad \nabla \phi(\bx) \cdot \hat{\boldsymbol{n}}\Big|_{\Gamma_{z_1}}= 0,\;\;
\int_{\Omega_p} \phi(\bx) \, d\bx = 0.
\end{aligned}
\right.
\label{eq:governing_equations CEF}
\end{equation}
The inclusion of $V_{\text{app}}(\bx)$ requires the energy functional in Equation~\eqref{eq: KS DFT variational problem} to be modified as:
\begin{multline}
      E_{\text{GS}}\left[ \left\{ \psi_n \right\},\left\{\bR^a \right\} \right] = \min_{\{\psi_i\}\in}{\max_{\phi\in}{\bigg\{T_s[\left\{ \psi_n \right\}] -\frac{1}{8\pi}\int_{\Omega_p}{|\nabla \phi(\bx)|^2d\bx} + \int_{\Omega_p}{(\rho(\bx)+b(\bx))\phi(\bx)d\bx}}} \\
    {{+\int_{\Omega_p}{(\rho(\bx)+b(\bx))V_\text{app}(\bx)d\bx}+E_{\text{xc}}[\rho(\bx)]+ E_{\text{loc}}[\rho(\bx)] + E_{\text{nloc}}[\{\psi_n\},\{\bR^a\}] \bigg\} }}
    \label{eq: KS DFT variational problem cef}  
\end{multline}
Additionally, the ionic forces are modified as
\begin{equation}
  \bF^{a} = \bF^a_{\text{loc}} + \bF^a_{\text{nloc}} + \bF^{a^*}_{\text{nloc}} + \bF^a_{\text{app}}   \label{eq: ion forces cef}
\end{equation}
where $\bF^a_{\text{loc}}, \bF^a_{\text{nloc}}$ are defined in Equation~\eqref{eq: Force expression} and $\bF^a_{\text{app}} = -\nabla V_{\text{app}}\mathcal{Z}_v$, with $\mathcal{Z}^a_v$ being the valence charge of atom '$a$'.

\subsubsection{\mtB~setup}\label{sec: constrained potential difference}
\begin{figure}[htp]
\centering
    \includegraphics[width=\textwidth]{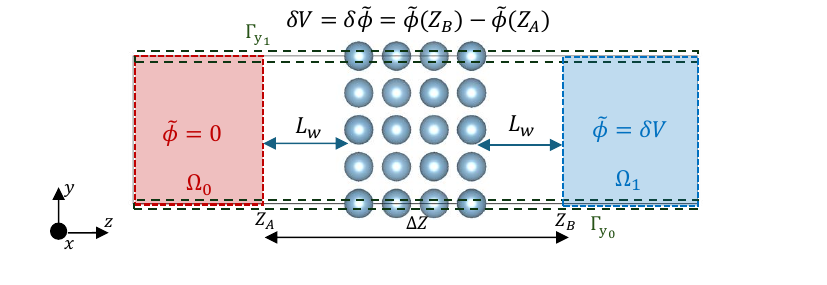}
        \caption{\textbf{\mtB}$:$ The boundary conditions for the total electrostatic potential ($\widetilde{\phi}(\bx)$) is periodic along the $x$ and $y$ directions. $\Gamma_{y_0}$, $\Gamma_{y1}$, $\Gamma_{x_0}$(not shown), and $\Gamma_{x_1}$(not shown) denote the periodic boundaries. $Z_A$ and $Z_B$ denote the interface between the vacuum and metal conductors. $L_w$ denotes the distance between the surface and metal conductors.
        Furthermore, $\widetilde{\phi}(\bx)$ is constrained in the region of metal conductors as shown in the shaded region, ensuring the electric field in the conductor region is zero.}
        \label{fig:APD}
\end{figure}
The natural approach to applying an external potential bias is to impose constraints on the total electrostatic potential ($\widetilde{\phi}(\bx)$) so that the desired potential difference is maintained. In contrast to the \mtA~setup, where  $\delta V_{\text{app}}$ is controlled via the value of the constant electric field, this method directly enforces the total electrostatic potential difference across the slab, $\delta \widetilde{\phi}$ as the boundary condition. Since $\delta \widetilde{\phi}$ corresponds directly to the controlling parameters used in electrochemical and surface science measurements, this approach provides a more direct link to the experimental setup.
 \begin{figure}[H]
\centering
    \includegraphics[width=\textwidth]{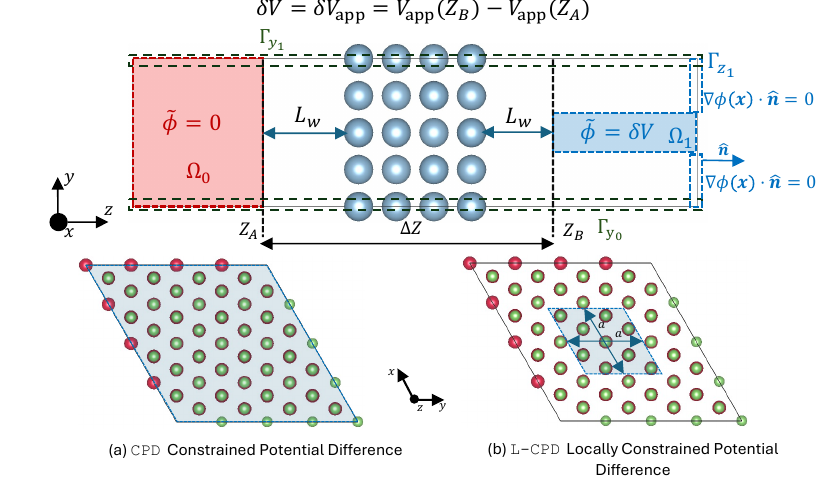}
        \caption{\cb \textbf{Top panel.} \textbf{\LmtB}$:$ The boundary conditions for the total electrostatic potential ($\widetilde{\phi}(\bx)$) is periodic along the periodic boundaries $\Gamma_{y_0}$, $\Gamma_{y1}$, $\Gamma_{x_0}$(not shown), and $\Gamma_{x_1}$(not shown). $Z_A$ and $Z_B$ denote the interface between vacuum and metal conductors. $L_w$ denotes the distance between the surface and metal conductors. $\widetilde{\phi}(\bx)$ is constrained in the region of metal conductors as shown in the shaded region, ensuring the electric field in the conductor is zero. \textbf{Bottom panel.} Comparison of (a) \mtB~and (b) \LmtB~setups when viewed along the $z$ axis for GaAs (Ga: red, As: green). The region in blue depicts the region on the surface where $\tilde{\phi}$ is constrained to be $V_{\text{app}}$. `$a$' indicates the length scale of the region where $\tilde{\phi}$ is applied in the \LmtB~setup.\cn }
        \label{fig:LCPD}
\end{figure}
To implement \mtB, the solution of the electrostatic variational problem in Equation~\eqref{eq: electrostatics Energy} should satisfy the boundary conditions shown in Figure~\ref{fig:APD}. \cb Furthermore, the \mtB~can be modified to localise the $\Omega_1$ region as shown in the top panel of Figure~\ref{fig:LCPD}, which we refer to as locally constrained potential difference (\LmtB) setup. The bottom panel of Figure~\ref{fig:LCPD} highlights the difference in the $\Omega_1$ region between the two setups\cn. Finally, we can summarize the governing equation and boundary conditions for computing the electrostatic potential ($\phi(\bx)$) in the  \mtB~\cb and the \LmtB~\cn setup as:
\begin{equation}
\left\{
\begin{aligned}
&-\frac{1}{4\pi}\nabla^2 \phi(\bx) = \rho(\bx) + b(\bx), \quad \bx \in \Omega_p,\\[1ex]
&\phi(\bx)\Big|_{\Gamma_{x_0}} = \phi(\bx)\Big|_{\Gamma_{x_1}}, \quad \phi(\bx)\Big|_{\Gamma_{y_0}}= \phi(\bx)\Big|_{\Gamma_{y_1}},\\[1ex]
&\phi(\bx) = -(V_{\text{loc}}(\bx)-V_{\text{sm}}(\bx))\; \forall \bx \in \Omega_0  \quad \phi(\bx) = \delta V -(V_{\text{loc}}(\bx)-V_{\text{sm}}(\bx))\; \forall \bx \in \Omega_1.
\end{aligned}
\right.
\label{eq:governing_equations APD}
\end{equation}

Notably, the Kohn-Sham energy functional remains unchanged from Equation~\eqref{eq: KS DFT variational problem}, thereby, the governing equation and the expression for ionic forces are exactly the same as in Equation~\eqref{eq: Governing equations} and ~\eqref{eq: Force expression}, respectively. We note that the electrostatic screening method,~\cite{Otani_ESM} designed for use with plane-wave codes is in a similar spirit. However, it relies on Green’s functions of the Poisson equation for various boundary conditions, restricting the applicability to cases where analytical solutions are available.

\subsection{FE discretization}
We discretize the governing equation in Equation~\eqref{eq: Governing equations} by employing the FE basis\cite{hughes2012finite,bathe}, which comprises of $\mathcal{C}^0$-continuous piecewise Lagrange polynomials interpolated over Gauss-Lobatto-Legandre nodal points. To this end, the FE representation of the various electronic fields in Equation~\eqref{eq: Governing equations} are given by,
\begin{equation}
    u^h_{n\bk}(\bx) = \sum_I^{M}{N^{h,p}_I(\bx)u^I_{n\bk}},\;\; \phi^h(\bx) =  \sum_I^{M_{el}}{N^{h,p_{el}}_I(\bx)\phi^I}
\end{equation}
where $u^I_{n\bk}$, $\phi^I$ present the FE discretized fields, while $N^{h,p}_I(\bx),N^{h,p_{el}}_I(\bx)$ are the strictly local Lagrange polynomials of degrees $p$, $p_{el}$, respectively. The resulting discretized eigenvalue problem $\bH^{\bk}\bu_{n\bk} = \varepsilon^h_{n,\bk}\bM\bu_{n\bk}$ is a nonlinear generalized eigenvalue problem where $\bH^{\bk}$ is the FE-discretised Hamiltonian and $\bM$ represents the FE-basis overlap matrix. Furthermore, to determine the electrostatics potential ($\phi^h(\bx)$), the FE-discretized Poisson equation $\bK \bphi = \bc $ is solved with appropriate boundary conditions, where the entries of $\bK$ are $K_{IJ} = \int_{\Omega_p}{\nabla N_I^{h,p_{el}}(\bx)\cdot \nabla N_J^{h,p_{el}}(\bx) d\bx}$ and $c_I = \int_{\Omega_p}{N_I^{h,p_{el}}(\bx)(\rho(\bx)+b(\bx))d\bx}$. A detailed discussion on the eigensolver, self-consistent-field iteration related mixing strategies and the efficient solution strategies that leverage the sparsity of the FE basis are discussed in our previous works\cite{dftfe1.0,dftfe0.6,JPDC, Das2022AcceleratingMatrix}.

\subsection{Computational details}
The two setups for applying an external potential bias (\mtA~and~\mtB), as discussed in the previous subsections, were implemented within the \DFTFE~framework employing norm-conserving pseudopotentials. In our calculations, we employed GGA\cite{GGA,GGAPerdew96} for the exchange-correlation functional, specifically utilizing the Perdew–Burke–Ernzerhof (PBE) form\cite{PBE}, as implemented in the \texttt{libxc}\cite{Libxc} library.  Furthermore, the plane-wave calculations for some of the validation studies were performed using \QE, using the \texttt{dipfield} option to enable dipole correction and the \texttt{tefield} option to add the sawtooth potential. The ONCV pseudopotentials\cite{oncv2013} used for these simulations were from pseudo-dojo\cite{PseudoDOjoONCV} and SPMS\cite{spms} repositories (refer to Supporting Information, Section S1 for more details). The plane-wave discretization parameter, $E_\text{cut}$ for \QE, was selected such that the discretization error with respect to refined calculation $(E_{\text{cut}}=100~E_h)$ is of $\order{10^{-5}}\frac{E_h}{\text{atom}}$ for DFT ground-state  energy, $\order{10^{-5}}\frac{E_h}{\text{bohr}}$ for ionic forces and $\order{10^{-6}}\frac{E_h}{\text{bohr}^3}$ for unit-cell stresses, wherever applicable. Similarly, the discretization parameters in \DFTFE~ are the FE interpolating polynomial degree '$p$' and mesh size around atom '$h$'. These were chosen such that a discretization error of $\order{10^{-5}}\frac{E_h}{\text{atom}}$ for ground-state energy, $\order{10^{-5}}\frac{E_h}{\text{bohr}}$ for ionic forces and $\order{10^{-6}}\frac{E_h}{\text{bohr}^3}$ was achieved with reference to a refined calculation $(E_{\text{cut}}=100~E_h)$ in \QE~for each bulk system. For the Brillouin zone integration, we employ Monkhorst-Pack (MP) grids\cite{PhysRevB.13.5188} to ensure systematic convergence of electronic properties. The $k$-point sampling rule for Brillouin zone integration was chosen so that the errors from successive refined samplings were of a higher order relative to the discretization errors incurred, ensuring systematic convergence of electronic properties. 

%% file: results.tex
In this section, we begin by benchmarking the \mtA~setup implemented in \DFTFE~against an equivalent setup\cite{QEdipoleCorrection,SawToothResta} in \QE. We compare the dipole moment, free energy, and ionic forces with various magnitudes of constant external electric fields and compute the dielectric response of the three systems considered in this work. Following this validation study for the \mtA~setup, we compare the \mtA~and the \mtB~setups by analyzing the difference in the ground state solutions of electron density ($\rho(\bx)$) and bare potential ($V_{\text{bare}}(\bx)$). Finally, we compare the influence of external potential bias in the two setups on ground-state properties, namely, surface and adsorption energies.
\begin{figure}[H]
    \centering
    \begin{subfigure}[b]{0.9\textwidth}
        \centering
\includegraphics[scale=1.0]{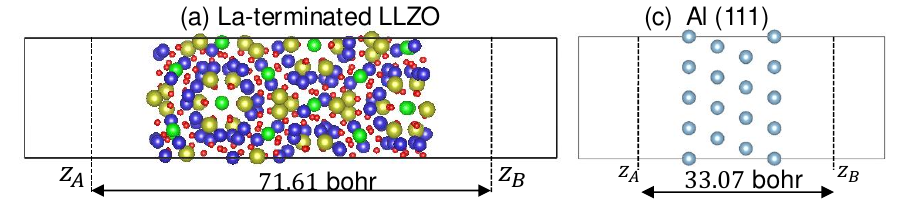} 
    \end{subfigure} 
    \vspace{1em} 
    \begin{subfigure}[b]{0.9\textwidth}
        \centering
\includegraphics[scale=1.0]{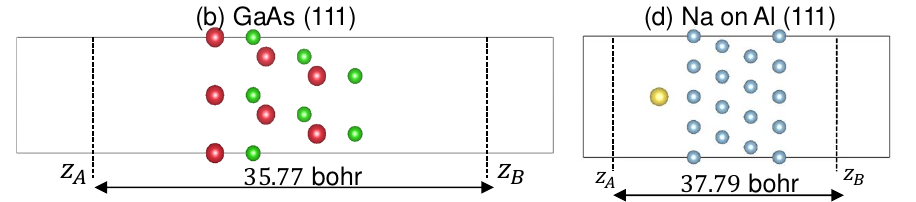}
    \end{subfigure}
\caption{Systems considered$-$ (a) La-terminated LLZO (Li: blue spheres, La: yellow, Zr: green, O: red), (b) GaAs (111) (Ga: red, As: green), (c) Al (111), and (d) Na adsorbed on Al (111) (Al: blue, Na: yellow). The locations $Z_A$ and $Z_B$ denote the metal-vacuum interfaces for the \mtB~setup. Furthermore, the external potential bias $\delta V$ across $Z_B$ -$Z_A$ is $\delta V_{\text{app}}$ in the \mtA~setup and $\delta \widetilde{\phi}$ in the \mtB~setup. }
\label{fig: Slab models}
\end{figure} 
The systems considered in this work, as showcased in Figure~{\ref{fig: Slab models}}, are: \textbf{(a) La-terminated LLZO slab}$-$ identified as one of the favourable terminations in a previous study \cite{LLZOMorphology}, consists of 12 formula units of Li\textsubscript{7}La\textsubscript{3}Zr\textsubscript{2}O\textsubscript{12}. LLZO is an insulating system,  has applications as a solid electrolyte in lithium-based batteries, and we employ $\Gamma$-point sampling for Brillouin zone integration.  \textbf{(b)  GaAs (111) slab}$-$ consists of four alternating layers of Ga and As atoms, comprising 24 atoms. The slab is polar, with Ga and As terminations on opposite surfaces. The bulk crystal structure was obtained from the Materials Project database, \cite{Jain2013} and is well-known to be a semiconducting material with applications in electronic devices. For Brillouin zone integration, we use a $10\times10\times1$ Monkhorst-Pack grid. \textbf{(c)  Al (111) slab} $-$ comprising four layers of the FCC structure with a total of 32 Al atoms. We used the bulk geometry of this metallic system from the Materials Project database \cite{Jain2013}. We use a $12\times12\times1$ k-point grid for Brillouin zone integration. As a sample system to model a simple adsorption process, we evaluate the adsorption energy of Na on Al (111). Note that we place the adsorbed Na at a distance of 5~bohr `above' the `top' layer of Al atoms in the (111) slab (see panel d in Figure~{\ref{fig: Slab models}}). We limit the maximum external electric field to $0.2\, \nicefrac{\text{V}}{\text{Å}}$ for the LLZO slab and $0.15\, \nicefrac{\text{V}}{\text{Å}}$ for the GaAs slab, where these limits are determined based on bulk calculated band gap and the slab thickness to ensure there is no dielectric breakdown.

\subsection{Validation of \mtA~setup}
In this subsection, we benchmark the \mtA~setup implemented in \DFTFE~as described in Section~\ref{sec: constant electric field} with that of the constant electric field setup\cite{SawToothResta,QEdipoleCorrection} used in \QE. For various magnitudes of external electric field $\mathcal{E}$, we compare the DFT internal energy, ionic forces, and dipole moments between the two codes. The dipole moment($\mu$) is computed relative to the center of the simulation domain as $\mu_z = \int_{\Omega_p}{ (b(\bx)+\rho(\bx))z d\bx}$, where $z$ represents the position along the non-periodic axis. We follow the convention that electron density ($\rho(\bx)$) is positive while the nuclear charge density ($b(\bx)$) is negative. The internal energy and forces are computed as per Equation~\eqref{eq: KS DFT variational problem cef} and Equation~\eqref{eq: ion forces cef}, respectively. 

As discussed in Section~\ref{sec: constant electric field}, referring to Equation~\eqref{eq:governing_equations CEF}, the electrostatic problem for the total charge density in the \mtA~setup is solved using Neumann boundary conditions on the non-periodic boundaries, with an additional zero-mean value constraint to fix the electrostatic potential reference. $\mathcal{E}$ is included in the DFT Hamiltonian using the auxiliary potential, $V_{\text{app}}$. 
In contrast, periodic boundary conditions are used in \QE~when computing the electrostatic potential. However, non-zero dipole moment in the system results in artificial electric fields in \QE~and is mitigated using a dipole correction scheme\cite{QEdipoleCorrection} in the course of the self-consistent-field iteration. The potential corresponding to the constant external electric field is of sawtooth form in both \DFTFE~and \QE, with maximum and minimum values positioned at 0.1 fractional units from the simulation boundaries. 
 
\begin{figure}[H]
    \centering
    \includegraphics[scale=0.95]{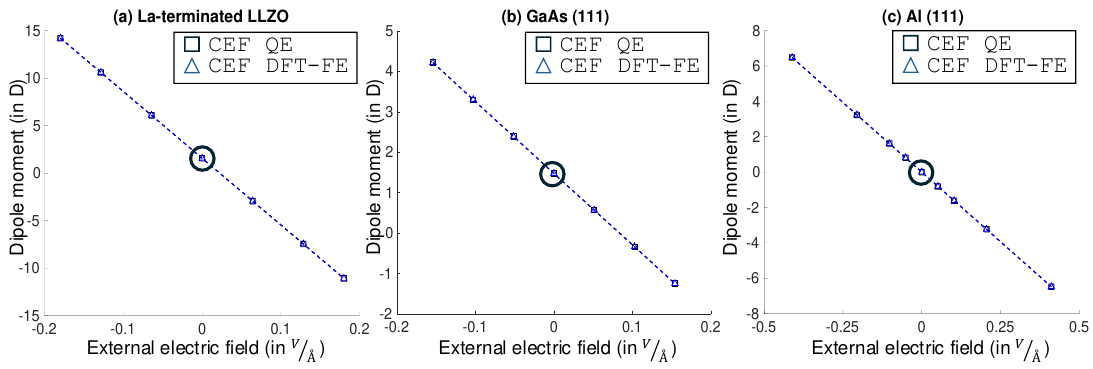}
    \caption{Dielectric response comparison of \mtA~in \DFTFE~(triangles)~against sawtooth potential with dipole correction in \QE~(squares). The plots show the dipole moment ($\mu$) in units of debye (D) as a function of applied external field ($\mathcal{E}$) in $\nicefrac{\text{V}}{\text{Å}}$ for (a) La-terminated LLZO, (b) GaAs (111), and (c) Al (111). The plots also highlight the dipole moment at zero external electric field. }  
    \label{fig: Dielectric response}
\end{figure}

Figure~\ref{fig: Dielectric response} compares the dielectric response of \mtA~implemented in \DFTFE, against the setup in \QE~for (a) La-terminated LLZO, (b) GaAs (111), and (c)  Al (111) slabs. The slope of the dipole moment against the external electric field quantifies the material's polarization ($\alpha_z = \frac{\Delta \mu}{\Delta \mathcal{E}}$). Table~\ref{tab: Results part 1} compiles the calculated dipole moments ($\mu$) for all systems at zero external electric field ($\mathcal{E}=0$) with \DFTFE~and \QE. For the  Al (111) slab, the presence of inversion symmetry results in a nearly vanishing dipole moment at $\mathcal{E} = 0$. In contrast, for the La-terminated LLZO and  GaAs (111) slabs, a nonzero dipole moment of 1.57 D and 1.47 D, respectively, is observed. These non-zero dipole moments highlight the importance of boundary conditions imposed: the presence of a large non-zero dipole moment suggests that if short circuit boundary conditions (i.e., periodic or homogeneous Dirichlet) were imposed on the total electrostatic potential ($\widetilde{\phi}(\bx)$), as would be the case in the \mtB~setup, the resulting ground-state obtained would be different. Hence, we emphasize that by imposing zero-Neumann boundary conditions (or open circuit boundary conditions) for slabs, we ensure that the electrostatic potential can adjust naturally, leading to a physically accurate description of the ground state. 

Furthermore, from Tables 2, 3 and 4 in Section~S2 of the Supporting Information, we observe a difference of $\order{10^{-6}}\frac{E_h}{\text{atom}}$ in free energy,  $\order{10^{-5}}\frac{E_h}{\text{bohr}}$ in ionic forces and $\order{10^{-3}}$D in dipole moment between \DFTFE~and~\QE. The close agreement in ground-state properties leads to excellent consistency in polarizability ($\alpha_z$) between \DFTFE~and \QE, as shown in Table~\ref{tab: Results part 1}, suggesting the equivalency of both approaches in implementing a \mtA. Having validated the \mtA~setup, we next compare this setup with the \mtB~setup, a more natural way of applying an external potential difference. We consider the same representative benchmark systems for the study. 

\begin{table}
    \centering
    
    \begin{tabular}{|c|c|c|c|c|}
     \hline
        \multirow{2}{*}{System} &  \multicolumn{2}{c|}{$\alpha_z$(in $\text{bohr}^3$)}   &  \multicolumn{2}{c|}{$\mu$ at $\mathcal{E}=0$ (in D)} \\
         \cline{2-5}
         & \QE &\DFTFE  & \QE  & \DFTFE \\
         \hline
         La-terminated LLZO& 1422.19 & 1422.27 & 1.5764  & 1.5756\\
         \hline
        GaAs (111)  & 357.89 & 357.92  & 1.4689  & 1.4713\\
        \hline
        Al (111)  & 318.56 & 318.62 & $1.95\times10^{-5} $& $1.98\times10^{-5} $\\
         \hline
    \end{tabular}
    \caption{Comparison of polarization ($\alpha_z$) and dipole moment ($\mu$) at zero external electric field ($\mathcal{E}=0.0$) for the benchmark systems.}
    \label{tab: Results part 1}
\end{table}

\subsection{Comparison between \mtA~and \mtB~setups}\label{sec: planar average comparison} 
In this section, we compare \mtA~and \mtB~setups of applying an external potential bias by analyzing planar average electron density (${\rho}^0(z)$) and planar average bare potential (${V}^0_{\text{bare}}(z)$), where $V_{\text{bare}}(\bx)$ is defined as
    $V_{\text{bare}}(\bx) = \widetilde{\phi}(\bx) + V_\text{app}(\bx)$ with
$\widetilde{\phi}(\bx) = \phi(\bx) + (V_{\text{loc}}(\bx) - V_{\text{self}}(\bx) )$ denoting the total electrostatic potential due to electron and nuclear charge density. Specifically, ${\rho}^0(z)$ and ${V}^0_{\text{bare}}(z)$ are computed as follows,
\begin{equation}
  {\rho}^{0}(z) = \int_{S_z}{\rho(\bx)dxdy},\;\; {V}^{0}_{\text{bare}}(z) = \frac{1}{A_z}\int_{S_z}{V_{\text{bare}}(\bx)dxdy} 
\end{equation}
where $S_z$ denotes the planar surface within the simulation domain located at position $z$ along the non-periodic axis, while $A_z$ corresponds to the area of $S_z$. Note that $V_\text{app}(\bx)$ denotes the applied potential arising due to the constant electric fields across the slab. As discussed in Section~\ref{sec: constant electric field}, \mtA~setup introduces $V_{\text{app}}$ as a sawtooth potential with slope $-\mathcal{E}$ across the slab. In contrast, the \mtB~setup (see Section~\ref{sec: constrained potential difference}) imposes the external potential bias as a constraint on the electrostatic potential ($\widetilde{\phi}$) at a distance of $L_w$ from the slab surface (see Figure~\ref{fig:APD}), while $V_{\text{app}}$ = 0 throughout the simulation domain. We position the metal-vacuum interface at $L_w=10$~bohr from the slab's surface to provide a sufficiently thick vacuum region that minimizes electron density penetration into the conductor, since the atomic valence density, as obtained from the pseudopotential file,\cite{spms,Pseudodojo} remains below 
$5\times10^{-6}\frac{e}{bohr^3}$ at $L_w=10$~bohr. 
Table~\ref{tab: Results part 2} shows the external potential bias imposed for the various systems considered, and this corresponds to an electric field of $\mathcal{E} = 0.1~\frac{V}{\text{Å}}$ for La-terminated LLZO and GaAs (111) and $\mathcal{E} = 0.15~\frac{V}{\text{Å}}$ for Al (111) with and without Na adsorbed in the \mtA~setup.

\begin{figure}[H]
    \centering
    \begin{subfigure}[b]{0.9\textwidth}
        \centering
        \includegraphics[scale=1.0]{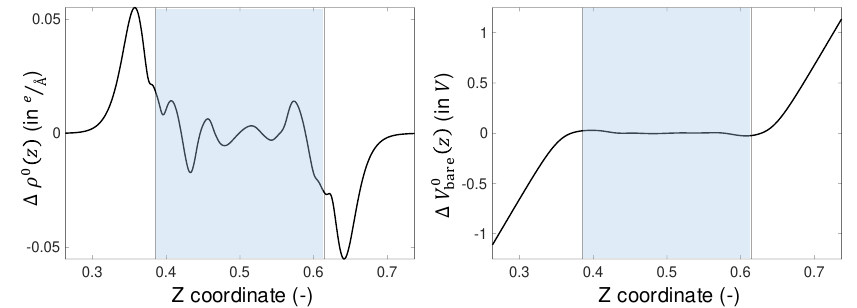}
        \caption{\textbf{GaAs (111)  slab} }
        \label{fig:GaAsPlanar}
    \end{subfigure}
    
    \vspace{1em} 

    \begin{subfigure}[b]{0.9\textwidth}
        \centering
        \includegraphics[scale=1.0]{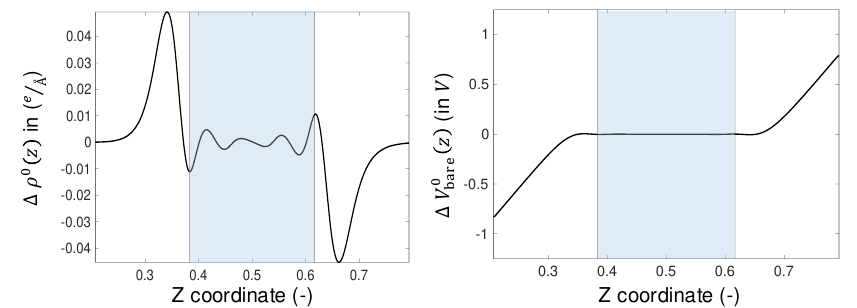}
        \caption{\textbf{Al (111) slab} }
        \label{fig:AlPlanarDifference}
    \end{subfigure}
    \vspace{1em} 

    \begin{subfigure}[b]{0.9\textwidth}
        \centering
        \includegraphics[scale=1.0]{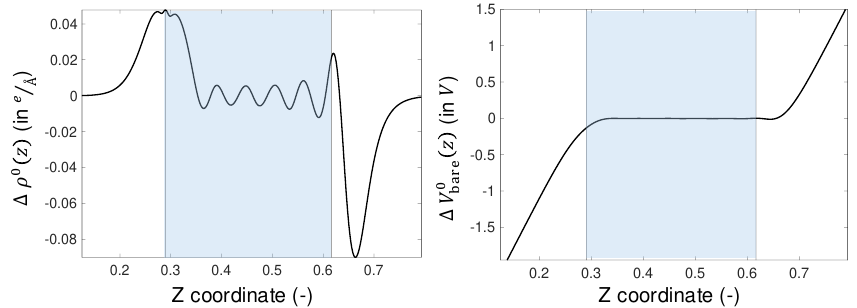}
        \caption{\textbf{Na adsorbed on  Al (111) FCC}}
        \label{fig:AlwithNaplanarDifference}
    \end{subfigure}

    \caption{Comparison of  \mtA~against \mtB. The shaded region depicts the region of the slab. The plot on the left shows $\Delta {\rho}^0(z)$ with respect to $z$-coordinate in fractional units, while the plot on the right shows $\Delta {V}^0_{\text{bare}}(z)$ with respect to $z$-coordinate in fractional units. We offset $V^0_{\text{bare}}(z)$ such that ${V}^0_{\text{bare}}(0.5) = 0$, aligning the reference for both the potentials.}
    \label{fig: Al and GaAs planar average}
\end{figure}

\begin{figure}[H]
\centering
    \includegraphics[scale=1.0]{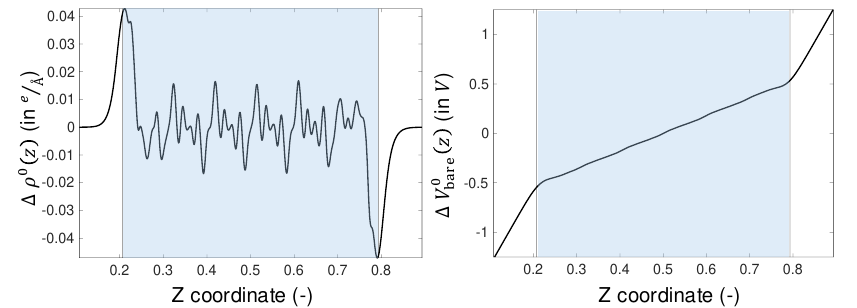}
    \caption{\textbf{La-terminated LLZO slab}. Comparison of  \mtA~against \mtB. The shaded region depicts the region of the slab. The plot on the left shows $\Delta {\rho}^0(z)$ with respect to $z$-coordinate in fractional units, while the plot on the right shows $\Delta {V}^0_{\text{bare}}(z)$ with respect to $z$-coordinate in fractional units. We offset $V^0_{\text{bare}}(z)$ such that ${V}^0_{\text{bare}}(0.5) = 0$, aligning the reference for both the potentials.}  
    \label{fig: LLZOplanarDifference}
\end{figure}

Figures~\ref{fig: Al and GaAs planar average} and \ref{fig: LLZOplanarDifference} show the plot of the difference in the ground-state planar average electron density ($\Delta \rho^{0}(z) = \rho^{0,\mtA}(z)- \rho^{0,\mtB}(z)$) and planar average bare potential ($\Delta V^{0}_{\text{bare}}(z) = {V}_{\text{bare}}^{0,\mtA}(z)-V_{\text{bare}}^{0,\mtB}(z)$). Additionally, Table~\ref{tab: Results part 2} compares the differences in dipole moments ($\mu$), free energies ($\Delta E$) and ionic forces ($\Delta F$) between the two setups studied in this work. For the case of Al (111) slab, with or without Na adsorbed, the intrinsic metallic screening results in negligible difference in $V_{\text{bare}}$ within the slab region between the two setups (Figure~{\ref{fig: Al and GaAs planar average}}), while a similarly strong internal screening is also observed in GaAs (111) slab, where no significant variation in $V_{\text{bare}}$ is observed within the slab between the two setups. In contrast, the insulating La-terminated LLZO slab shows a significant difference in $V_{\text{bare}}$ between the two setups within the slab region (Figure~{\ref{fig: LLZOplanarDifference}}). 
 
As discussed earlier in this subsection, the potential bias is applied such that the potential at $Z_A$ is higher than that of $Z_B$ (see Figures~\ref{fig:CEF} and \ref{fig:APD}). Consequently, we observe the electron density, and hence the dipole moment ($\mu$) to shift towards $Z_B$. Furthermore, $\mu$ values in Table~\ref{tab: Results part 2} indicate that the shifts in $\mu$ for the \mtB~setup is greater than \mtA, which is due to the fact that the \mtB~setup precisely maintains the target potential bias, while $\delta V^0_{\text{bare}}$ is lower in the \mtA~setup. The $\delta V^0_{\text{bare}}$ being lower than the target value in the \mtA~setup is expected since $\delta V_{\text{app}} (\neq \delta V^0_{\text{bare}})$ is obtained via the electric field $\mathcal{E}$, which is the controlling parameter. On the other hand, $\delta \widetilde{\phi}(=\delta V^0_{\text{bare}})$ is the controlling parameter in the \mtB~setup.  As a consequence, we treat the control parameters, namely $\mathcal{E}$ or $\delta V_{\text{app}}$ in \mtA~and $\delta V^0_{\text{bare}}$ or $\delta \widetilde{\phi}$ in \mtB, for each setup separately and emphasize that for experimental setups where the total electrostatic potential ($\widetilde{\phi}$) across the surface or interface is controlled, the \mtB~setup is a natural choice in the modelling strategy. In the next subsection, we present the surface energies and adsorption energies using both setups.
 
 

\begin{table}
    \centering
    
    \footnotesize{ \begin{tabular}{|c|c|c|c|c|c|c|c|}
     \hline
       \multirow{2}{*}{System} & \multirow{2}{*}{Target $\delta V$ (in $\frac{E_h}{e}$)} & \multicolumn{2}{c|}{$\delta V^0_{\text{bare}}$ (in $\frac{E_h}{e}$)}   &  \multicolumn{2}{c|}{$\mu$ (in D)}& \multirow{2}{*}{$\Delta$E (in $\frac{E_h}{\text{atom}}$)} & \multirow{2}{*}{$\Delta$F (in $\frac{E_h}{\text{bohr}}$)} \\
         \cline{3-6}
         & & \mtA &\mtB  & \mtA  & \mtB & & \\
         \hline
         La-terminated LLZO&-0.143& -0.072 & -0.143 & 8.2  & 19.39 & $3.51\times 10^{-5}$& $4.04\times10^{-3}$\\
         \hline
         GaAs (111)&-0.072 & -0.008 & -0.072   & 3.3  & 8.64 &$2.22\times10^{-4}$ & $7.78\times10{-4}$  \\
        \hline
        Al (111)&-0.099  & -0.040 & -0.099 & 2.43 & 6.08 & $1.42\times10^{-4}$ &$9.73\times10^{-5}$\\
         \hline
        Al (111) + Na&-0.113  & -0.021 & -0.113 & 5.5 & 13.83  &$3.81\times10^{-4}$ & $2.6\times10^{-3}$\\
         \hline
    \end{tabular}}
    \caption{Comparison of $\delta V^0_{\text{bare}}$ compared to the target value, dipole moments ($\mu$), differences in free energies ($\Delta E$) and ionic forces ($\Delta F$) between the \mtA~and \mtB~setups for the benchmark systems considered. }
    \label{tab: Results part 2}
\end{table}

\subsection{Surface and adsorption energies: \mtA~vs.~\mtB}
In this section, we examine the relaxed surface energies of the La-terminated LLZO surface and GaAs (111)  surface at various external potential biases. Furthermore, we also investigate the effect of external potential bias on the adsorption energy of Na on Al (111)  surface. We compare the surface energies ($\gamma$) and adsorption energies ($E_{\text{ads}}$) obtained between the two setups \mtA~and \mtB. 
All relaxed structures are obtained using the LBFGS algorithm until the atomic forces are below $4\times10^{-4}\frac{E_h}{\text{bohr}}$, ensuring well-converged structures for subsequent analyses. 
 
\paragraph{Comparison of surface energies:} The surface energy ($\gamma$) in $\nicefrac{J}{m^2}$ is computed from DFT as $\gamma(\delta V) = \frac{1}{2A}\bigg[ E_{\text{slab}}(\delta V)-\frac{N}{N_{\text{bulk}}}E_{\text{bulk}} \bigg]$, where $A$ denotes the surface area in $m^2$. $E_{\text{slab}}(\delta V)$ and $E_{\text{bulk}}$ are the DFT total energies of the relaxed slab and bulk structures, with $\frac{N}{N_{\text{bulk}}}$ indicating the ratio of the number of formula units present in the slab to the number of formula units in the bulk unit cell. In the \mtA~setup, the potential bias is such that $\delta V_{\text{app}} = \delta V$, while for \mtB~setup, the total electrostatic potential difference satisfies $\delta \widetilde{\phi} = \delta V$ (see Figures~\ref{fig:CEF} and~\ref{fig:APD}). 

\begin{figure}[htp]
    \begin{subfigure}[b]{\textwidth}
\includegraphics[scale=1.1]{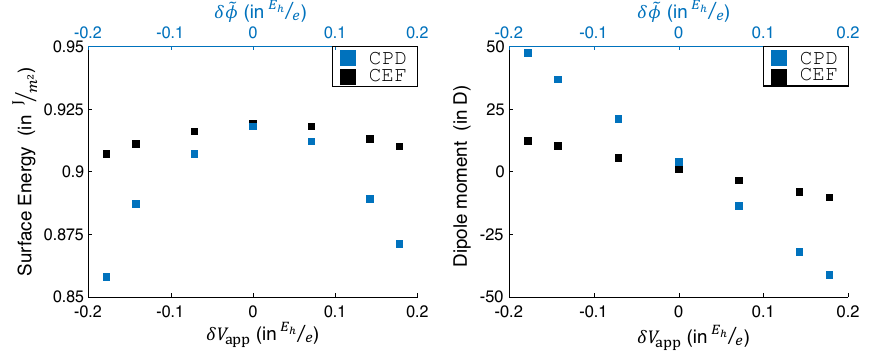}
    \caption{\textbf{Case Study: La-terminated LLZO slab}}  
    \label{fig: LLZO surface energy}
    \end{subfigure}
    
    \vspace{1em} 

    \begin{subfigure}[b]{\textwidth}
\includegraphics[scale=1.1]{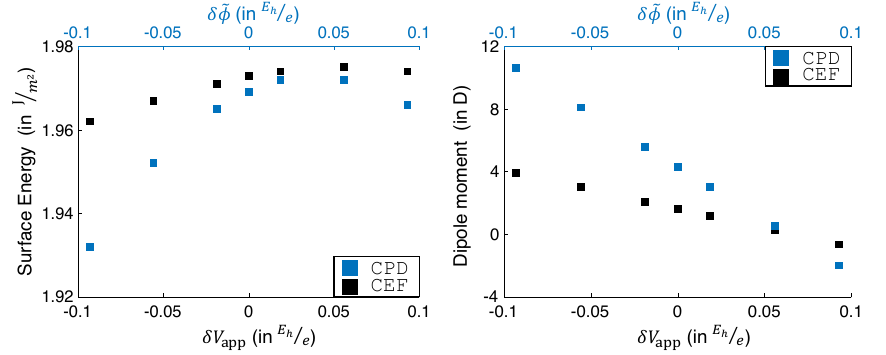}
    \caption{\textbf{Case Study: GaAs (111) slab}} 
    \label{fig: GaAs surface energy}
    \end{subfigure}
\caption{Comparison of surface energy (in $\nicefrac{J}{m^2}$) and dipole moment (in debye) between \mtA~and \mtB~setups. The systems considered are (a) \textbf{La-terminated LLZO} (top row) and (b) \textbf{GaAs (111) slab} (bottom row). The control parameters, $\delta V_{\text{app}}$ (black) for \mtA~and $\delta \widetilde{\phi}$ (blue) for \mtB~are shown as separate x-axes.}
\label{fig: surface energy}
\end{figure}

Figure~\ref{fig: surface energy} compares the surface energies ($\gamma$) and dipole moments ($\mu$) for La-terminated LLZO and  GaAs (111) slabs. Note that each panel in Figure~\ref{fig: surface energy} shows the control parameter $\delta V_{\text{app}}$ (bottom x-axis) for \mtA~and $\delta \widetilde{\phi}$ (top x-axis) for \mtB~separately. From Figure~\ref{fig: surface energy}(a) for La-terminated LLZO, we observe that the surface energy of \mtB~is consistently lower than \mtA. Moreover, the difference in surface energies increases with increasing magnitude of potential difference, with a maximum difference of 0.049~$\nicefrac{J}{m^2}$ at a potential difference ($\delta V_{\text{app}}$) of -0.178 \nicefrac{$E_h$}{$e$}. A similar trend is observed in Figure~\ref{fig: surface energy}(b) for GaAs (111)  slab, where the surface energy computed in \mtB~setup is consistently lower than \mtA, with the maximum difference in $\gamma$ of 0.03~$\nicefrac{J}{m^2}$. The minimum difference in $\gamma$ between the two setups occurs at a positive bias of 0.018 \nicefrac{$E_h$}{$e$} for GaAs (111), while the minimum difference in $\gamma$ for LLZO occurs at zero bias. Additionally, for both the LLZO and GaAs, we observe the variation of dipole moment ($\mu$) of relaxed structures in the \mtB~setup to exhibit a steeper variation with applied bias than \mtA. The steeper variation of $\mu$ in \mtB~is in line with our previous observation of a larger dipole shift in the \mtB~setup, as described in Section~\ref{sec: planar average comparison}. The differences observed in $\gamma$ and $\mu$ between the two setups indicate the fundamental differences in interpreting an applied potential bias across a system and the importance of the different control parameters involved.

\begin{figure}[htp]
\includegraphics[scale=1.1]{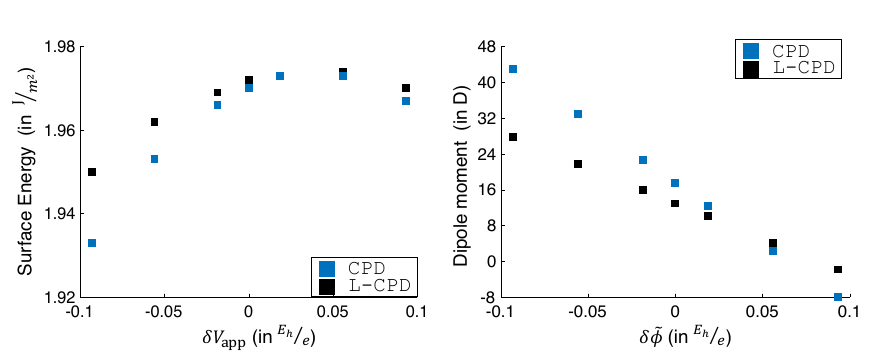}
\caption{\cb Comparison of surface energy (in $\nicefrac{J}{m^2}$) and dipole moment (in debye) between \mtB~and \LmtB~setups in a  \textbf{$2\times2\times1$ GaAs (111) slab}. Since a $2\times2\times1$ supercell is considered, the dipole moment is $4\times$ that in Figure~\ref{fig: GaAs surface energy}.\cn }
\label{fig: GaAs surface energy2}
\end{figure}

\cb
To demonstrate the differences in surface energies obtained with the \LmtB~setup against the \mtB~setup, we compare the surface energies of the $2\cross 2\cross 1$ supercell of the GaAs slab, as a function of the applied potential ($\delta \widetilde{\phi}$), as shown in Figure~\ref{fig: GaAs surface energy2}. Importantly, we observe that the \mtB~setup always has lower surface energy, and the variation of dipole moment is steeper in the \mtB~setup than the \LmtB~setup. These trends are consistent with the fact that the potential is applied over a larger surface area in the \mtB~setup compared to the \LmtB~approach that allows for a larger modification of the underlying electronic structure in \mtB~versus \LmtB, resulting in a lowering of the surface energy and a larger change in the dipole moment. Additionally, we observe that the trends in surface energy and dipole moment corresponding to \LmtB~setup lie between the \mtA~and \mtB~setups, indicating that the \LmtB~ adds a modification of the electronic structure that is intermediate between the two approaches.

\cn
\paragraph{Adsorption energy:} The adsorption energy ($E_{\text{ads}}$) of Na on Al (111) surface is computed as $E_{\text{ads}}(\delta V) = [ E^{\text{Al+Na}}_{\text{slab}}(\delta V) - E^{\text{Al}}_{\text{slab}}(\delta V) -E^{\text{Na}}]$, where $E^{\text{Al+Na}}_{\text{slab}}(\delta V)$ denotes the DFT internal energy of Na atom on Al (111) slab and $E^{\text{Al}}_{\text{slab}}(\delta V)$ denotes the internal energy of Al (111) slab at external potential bias $\delta V$. $E^{\text{Na}}$ is the internal energy of single Na atom in vacuum without any potential bias.  We ensure that the locations ($Z_A$ and $Z_B$) of the metal-vacuum interface for both Na on Al (111) slab and standalone Al (111) slab are the same with $\Delta Z = Z_A - Z_B = 33.8~\text{bohr}$.

\begin{figure}[htp]
\includegraphics[scale=1.1]{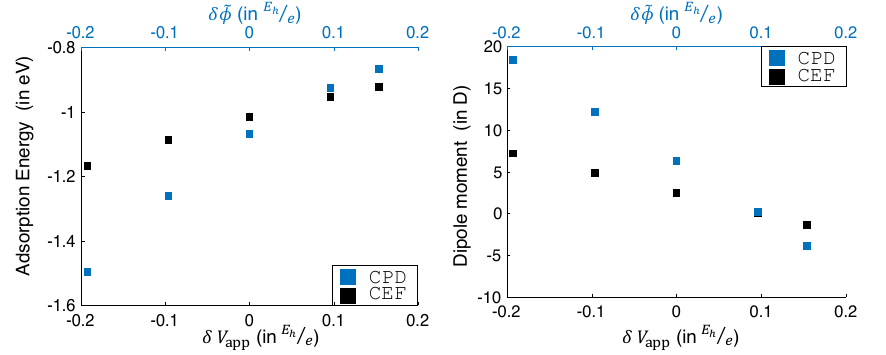}
    \label{fig: Al(111)-Na adsorption energy}
\caption{Comparison of adsorption energy (in eV) and dipole moment (in debye) between \mtA~and \mtB. The adsorption energy of \textbf{Na on Al (111) surface} is computed along with the dipole moment of the relaxed Al (111) slab at various external potential biases. The control parameters $\delta V_{\text{app}}$ (black) for \mtA~and $\delta \widetilde{\phi}$ (blue) for \mtB~are shown as separate x-axes.}
\label{fig: Adsorption energy}
\end{figure}

Figure~\ref{fig: Adsorption energy} compares the adsorption energy ($E_{\text{ads}}$) and dipole moments ($\mu$) of the relaxed structures of Na adsorbed on Al (111) and Al (111) slab for both the setups. Similar to the surface energy plots (Figure~\ref{fig: surface energy}), the panels in Figure~\ref{fig: Adsorption energy} show the control parameter $\delta V_{\text{app}}$ (bottom x-axis) for \mtA~and control parameter $\delta \widetilde{\phi}$ (top x-axis) for \mtB~separately. From Figure~\ref{fig: Adsorption energy}, we observe that the adsorption energy decreases with decrease in external potential bias for both setups, which can be attributed to Na being near the $Z_A$ interface and adsorbs on the Al by partial electron transfer (see Figure~\ref{fig: Slab models}). Under a positive bias, the electron density is drawn toward the Na atom, inhibiting further transfer of electrons and thereby weakening the ability of Na to get adsorbed. In contrast, a negative bias redistributes electron density away from the Na site, promoting charge transfer and strengthening Na adsorption. Additionally, from Figure~\ref{fig: Adsorption energy}, we observe that the change in Na adsorption energy is more sensitive to the applied bias in the \mtB~setup than the \mtA~setup. Similarly, we observe a steeper variation in the $\mu$ of relaxed structures versus applied bias in the \mtB~setup compared to \mtA, in agreement without our observation in Section~\ref{sec: planar average comparison}.

%% file: Discussion.tex
Accurate and efficient first-principles modelling of surfaces and interfaces is essential for gaining theoretical insights into essential processes such as charge transfer, reaction kinetics, material stability, and polarization, which are highly important to understand and optimize in applications such as catalysis, batteries, fuel cells, sensors, and electronics. While PW-DFT has long been the preferred method for first-principles simulations, the inherent restriction in plane-wave methodologies to impose periodic boundary conditions leads to undesirable consequences such as spurious image-image interactions and the emergence of artificial electric fields in the case of systems with a net-dipole. Moreover, plane-wave-based codes exhibit poor scalability on multi-node CPU-GPU architectures, restricting the system sizes that can be handled. In contrast, real-space FE methods employed in this work can accommodate generic boundary conditions and have demonstrated exceptional ability to scale on massively parallel supercomputing architectures across the world. The ability of \DFTFE~to efficiently handle large systems and accommodate generic boundary conditions presents new opportunities for modelling surfaces and interfaces with minimal approximations, which has been leveraged in this work for modelling surfaces and interfaces.


The control of external parameters, such as potential bias, solvation effects, or their combination, plays a critical role in tailoring the properties of slabs and interfaces. PW-DFT calculations using a constant electric field\cite{SawToothResta, PhysRevB.64.125403,PhysRevB.63.205426} have provided insights on controlling properties such as surface diffusion, polarization and ferroelectricity. Additionally, the effective screening medium (ESM) method\cite{Otani_ESM} in PW-DFT decouples the periodicity of the electrostatic potential by analytically solving for the electrostatic potential using the Greens function approach with non-periodic boundary conditions. This method provides flexibility in modelling surfaces and interfaces by introducing a generic framework to control potential bias, solvation effects or their combination. 
However, when employed in conjunction with PW-DFT, the ESM method assumes that the mean field effective potential is short-ranged, which is not necessarily true when  exact exchange\cite{b3lyp} or van-der-waals\cite{Dion2004} functionals are employed.  Consequently, employing semi-periodic boundary conditions is essential for the accurate modelling of surfaces and interfaces without any spurious periodic interactions in the presence of a potential bias. 

Addressing the above limitations and to model larger-scale systems involving surfaces and interfaces, we implement in \DFTFE~two setups of applying an external potential bias: (i) constant electric field (\mtA) and (ii) constrained potential difference (\mtB). These setups, in contrast to the NEGF (nonequilibrium Green’s functions)\cite{negf} formulation, ensure that the electrons are in the ground state and the electronic current is negligible. We benchmark and validate the \mtA~setup with the constant electric field setup in \QE~by comparing ground-state properties such as internal energy, ion forces and dipole moment. We observe an excellent agreement in the ground-state properties for the benchmark systems considered, namely, LLZO, GaAs, and Al. 

In the \mtA~setup, a constant electric field ($\mathcal{E}$) is applied along the non-periodic direction in \DFTFE. The DFT Hamiltonian is modified by introducing an auxiliary linear potential, $V_{\text{app}}(\bx)$, such that the slope of this linear potential equals $-\mathcal{E}$. Furthermore, the electrostatic potential arising from the electron and nuclear densities is obtained by solving the Poisson's problem with zero-Neumann boundary conditions on boundaries parallel to the slab surface. Additionally, a zero mean-value constraint is imposed to fix the reference of the electrostatic potential ($\phi(\bx)$). In this setup, the modelling of surfaces and interfaces in a vacuum can be accomplished by setting the external electric field to zero ($\mathcal{E}$=0). 

In contrast to the \mtA~setup, the 
\mtB~setup directly enforces the desired (experimental) potential bias by imposing constraints on the total electrostatic potential ($\widetilde{\phi}(\bx)$). \cb Note that the \mtB~setup can be modified to include solvation effects\cite{implicit1, implicit2} with appropriate changes to the Poisson problem to determine the total electrostatic potential.\cn~The two setups (\mtA~and \mtB) control different parameters ($\delta V_{\text{app}}$ and $\delta \widetilde{\phi}$, respectively) and hence have different electronic ground-states as demonstrated in Figures~{\ref{fig: Al and GaAs planar average}} and {\ref{fig: LLZOplanarDifference}}. \cb Additionally, the \LmtB~setup, an extension of the \mtB~setup is proposed by localizing the region where $\delta \widetilde{\phi}$ is controlled, which can have implications in modelling physical systems exposed to potential differences in localised regions. Note that the \LmtB~setup demonstrated in this work can not be trivially implemented in PW-DFT.\cn

To further contrast the two setups (\mtA~and \mtB), we compared the planar average electron density ($\rho^0(z)$) and bare potential within our representative systems. We observed for (111) Al FCC and (111) GaAs a similar behaviour of $V^0_{\text{bare}}(z)$ within the slab due to screening effects, while for La-terminated LLZO slab, we noticed a significant difference in $V^0_{\text{bare}}(z)$ in the region of the slab. For all the systems considered, we observed that the dipole moment response to the potential bias was stronger in the \mtB~setup than \mtA, resulting in steeper variations in surface and adsorption energies with applied bias. Overall, the \mtB~setup consistently demonstrates greater sensitivity to the external potential bias in both surface energy and adsorption energy calculations compared to the \mtA~setup, and should represent experimental scenarios of an applied potential bias better. Finally, the two setups (\mtA~and \mtB) as implemented in \DFTFE~offer a robust framework for investigating surfaces and interfaces without any underlying assumptions or correction schemes, while also enabling simulations at that scale better with available computational resources compared to most state-of-the-art PW-DFT codes. 



%% file: acknowledgement.tex
P.M. gratefully acknowledges the seed grant from the Indian Institute of Science and the SERB Startup Research Grant from the Department of Science and Technology India (Grant Number: SRG/2020/002194), which supported the purchase of GPU clusters used in the current work. P.M and S.G acknowledge National Supercomputing Mission (NSM) R\&D for exascale grant (DST/NSM/R\&D Exascale/2021/14.02) that supported the use of PARAM Pravega supercomputer at the Indian Institute of Science. This research also used computational resources at the Argonne and the Oak Ridge Leadership Computing Facilities. The Argonne Leadership Computing Facility at Argonne National Laboratory is supported by the Office of Science of the U.S. DOE under Contract No. DE-AC02-06CH11357. The Oak Ridge Leadership Computing Facility at the Oak Ridge National Laboratory is supported by the Office of Science of the U.S. DOE under Contract No. DE-AC05-00OR22725. A portion of the
calculations in this work used computational resources of the
supercomputer Fugaku provided by RIKEN through the HPCI
System Research Project. K.R. acknowledges the Prime Minister Research Fellowship (PMRF) from the Ministry of Education India for financial support. P.M. acknowledges the Google India Research Award 2023 for financial support during the course of this work.  